\documentclass[acmsmall]{acmart}

\AtBeginDocument{%
  }

\setcopyright{cc}
\setcctype{by-nd}
\acmJournal{PACMHCI}
\acmYear{2025} \acmVolume{9} \acmNumber{7} \acmArticle{CSCW411} \acmMonth{11} \acmPrice{}\acmDOI{10.1145/3757592}

\acmISBN{978-1-4503-XXXX-X/18/06}

\newcommand{\kedit}[1]{\textcolor{black} {#1}}

\usepackage{soul}
\begin{document}

\title{Exit Stories: Using Reddit Self-Disclosures to Understand Disengagement from  Problematic Communities}

\author{Shruti Phadke}
\email{shruti.phadke@drexel.edu}
\orcid{0000-0002-0524-830X}
%\email{webmaster@marysville-ohio.com}
\affiliation{%
 \institution{Drexel University}
 \city{Philadelphia}
 \state{PA}
 \country{USA}
}

\renewcommand{\shortauthors}{Shruti Phadke}

\begin{abstract}
  Online platforms like Reddit are increasingly becoming popular for individuals sharing personal experiences of leaving behind social, ideological, and political groups. Specifically, a series of ``ex-'' subreddits on Reddit allow users to recount their departures from commitments such as religious affiliations, manosphere communities, conspiracy theories or political beliefs, and lifestyle choices. Understanding the natural process through which users exit, especially from problematic groups such as conspiracy theory communities
  and the manosphere,  can provide valuable insights for designing interventions targeting disengagement from harmful ideologies. This paper presents an in-depth exploration of 15K \textit{exit stories} across 131 subreddits, focusing on five key areas: religion, manosphere, conspiracy theories, politics, and lifestyle. Using a transdisciplinary framework that incorporates theories from social psychology, organizational behavior, and violent extremism studies, this work identifies a range of factors contributing to disengagement. The results describe how disengagement from problematic groups, such as conspiracy theories and the manosphere, is a multi-faceted process that is qualitatively different than disengaging from more established social structures, such as religions or political ideologies. This research further highlights the need for moving beyond interventions that treat conspiracy theorizing solely as an information problem and contributes insights for future research focusing on offering mental health interventions and support in exit communities.
  \end{abstract}





\maketitle

\section{Introduction}
Increasingly, internet users are turning to online communities to share their experiences of leaving behind past social affiliations or groups. Platforms like Reddit have become a popular choice for individuals to connect around shared experiences, particularly when leaving ideological or social commitments behind. Reddit hosts a growing culture of ``ex'' communities, such as \texttt{r/exmormon}, \texttt{r/ReQovery} or \texttt{r/exredpill}, where users post detailed personal narratives---\textit{exit stories}---capturing the users' self-disclosures around disengaging or exiting from religious, political, ideological communities or lifestyle commitments (see Figure \ref{fig:rq1_examples} for example). These \textit{exit stories} are particularly significant for understanding how people naturally disengage from problematic ideologies, such as conspiracy theories, manosphere beliefs, or extremist political movements. 

I build on the previous qualitative studies providing valuable insights into why people leave social affiliations \cite{ebaugh1988becoming}, religions \cite{coates2013disaffiliation}, conspiracy theory beliefs \cite{engel2023learning,xiao2021sensemaking}  or manosphere groups \cite{thorburn2023exiting}, and contribute a data-driven understanding of different factors influencing users' disengagement. In this paper, I use the term disengagement to refer to the self-reported process of distancing oneself—cognitively, emotionally, or socially—from a particular belief system, identity group, or lifestyle. While the data originates from online spaces, the disengagements described often reflect broader offline identity shifts and social consequences. Disengagement may involve leaving a single community, distancing from multiple related groups (e.g., red pill and MGTOW), or rejecting the underlying ideology altogether. The scope and completeness of disengagement vary by story, and this study focuses on capturing the self-perceived moments and motivations that users themselves frame as ``exiting``.

%Specifically, in this paper, I take an exploratory dive into Reddit's exit communities and ask:
Studying voluntary exit—especially through users’ own narratives—offers critical insights into belief change, identity transformation, and the role of online infrastructures in supporting or hindering recovery. However, detecting and analyzing these "exit stories" at scale is methodologically challenging: such disclosures are not always explicitly labeled or uniformly structured, and they appear across a wide range of communities. This motivates the first research question, which is exploratory and procedural in nature:

\vspace{5pt}

\noindent \textbf{RQ1: } What types of \textit{exit stories} are shared on Reddit? 
\vspace{5pt}

To answer RQ1, I adopt a data-driven approach, systematically identifying 131 subreddits where \textit{exit stories} are shared, covering five key thematic areas: Religion, Manosphere, Conspiracy Theories, Politics, and Lifestyle. Along with the posts shared in these subreddits, I also leverage Reddit's Q\&A culture, both inside and outside of the 131 subreddits. The \textit{exit stories} dataset provides a robust quantitative foundation for investigating the intricacies of disengagement across various themes. 

Once \textit{exit stories} are identified, the next step is to understand the underlying factors in disengagement. This is especially important towards supporting emerging research that is increasingly focusing on designing platforms and interventions that mitigate online harm and support user well-being. Prior research has often treated ideological disengagement as a cognitive or informational correction, such as countering misinformation, but much less is known about how individuals naturally come to reject problematic communities, on their own terms, in their own words. Hence, the second research question asks: 

\vspace{5pt}
\noindent \textbf{RQ2: }What are the various factors motivating the process of disengagement shared in \textit{exit stories}?

\vspace{5pt}

To understand various factors in disengagement, in RQ2, I employ a transdisciplinary approach, rooted in theories on disengagement from social psychology \cite{tajfel1978intergroup,becker1963outsiders,ebaugh1988becoming,festinger1954theory}, organizational exit \cite{hirschman1970exit,hom2011organizational}, violent extremism \cite{aho1988out,morrison2021systematic}, and human-computer interaction \cite{engel2023learning,xiao2021sensemaking,thorburn2023exiting}. I develop a comprehensive theory-driven \textit{exit reasons framework} which is further refined through qualitative coding. To derive insights from the \textit{exit stories}  on a large scale, I further apply the \textit{exit reasons framework} using a GPT-4 model in a multi-label classification task.

The results reveal distinct patterns of disengagement across different themes.
Overall, conspiracy theory and political exits are largely influenced by the presence of contradictory exposure. However, disengagement from conspiracy theories is also prominently driven by mental health and moral reasons. Manosphere \textit{exit stories} overwhelmingly mention negative self evaluation as one of the reasons for disengagement whereas Lifestyle \textit{exit stories} largely discuss negative physical health. Supporting the observations from the disaffiliation literature \cite{sandomirsky1990processes}, religious exits are primarily attributed to coercive or strict control and moral conflicts.

% The results from t
\kedit{T}his paper provide\kedit{s} valuable insight\kedit{s} for intervention design, highlighting the need to move beyond information-centric approaches for countering problematic worldviews. I further outline how results from this work can motivate future research directions in building accessible mental health support into the designs of exit communities. By comparing disengagement reasons across various types of social groups, I outline how disengaging from conspiracy theory or manosphere groups is qualitatively different than leaving behind religion, lifestyle commitments, or political ideologies. These findings also support the notion that disengagement is not a singular event but a process involving several cognitive, psychological, moral, and identity factors.

Specifically, this work makes the following contributions:

\begin{itemize}
    \item 
    % This paper offers a 
    \kedit{A} comparative, data-driven understanding of 
    % factors in 
    disengagement from five different types of groups---religious, political, conspiracy theory, manosphere, and lifestyle. 
    \item 
    % This work contributes a 
    \kedit{A} transdisciplinary\kedit{,} theory-driven, and data validated \textit{exit reasons framework} that outlines social, emotional, cognitive and other factors influencing the general process of disengagement. 
    \item 
    % This work provides an e
    \kedit{E}mpirical evidence that cognitive and information factors are not the sole drivers in disengaging from problematic beliefs and argues for the need of more holistic intervention approaches.
\end{itemize}

This work contributes to the CSCW and HCI literature by offering a large-scale, theory-driven investigation into how people disengage from various social affiliations and how they disclose their exit process in online communities. 

Prior CSCW and social computing research has examined how online communities support identity exploration, behavior change, and peer support (e.g., in mental health {\cite{de2014mental}}, addiction recovery {\cite{rubya2017video}}, or quitting smoking {\cite{de2023helping}}), but has less frequently examined naturalistic, voluntary disengagement from ideologically-bound communities such as conspiracy theorists or manosphere groups. By surfacing exit stories shared voluntarily by users, this work can demonstrate how platform infrastructure (e.g., Reddit’s subreddit design and Q\&A culture) enables identity work and reflection during disengagement. The insights from this work can expand CSCW’s focus from community engagement to community disengagement, and from health or productivity-related goals to ideological and identity-oriented change.

\section{Background}
To investigate \emph{why} users exit from their communities, I draw on both current and foundational work from social psychology, sociology, computer-supported cooperative work, and ideological domain studies. I further outline traditional and emerging methods for extracting cause or reasoning from social media data. 

\subsection{\textbf{Research on disengagement from various perspectives}}

\subsubsection{\textbf{The social-psychological perspective on disengagement}}
\label{lit:soc-psych}
Social psychology offers two distinct but complementary outlooks on the process of disengagement. The social identity scholarship \cite{tajfel1978intergroup}  describes how individuals define themselves through their membership in social groups. It emphasizes the psychological attachment to group identities and the cognitive processes that drive group behavior, including group membership and commitment. While most of the literature is focused on understanding affirmative aspects of group membership and commitment, i.e, what makes people commit to the groups, several studies also outline how people may leave groups due to misaligned identities \cite{hirsh2016mechanisms} and morals\cite{monin2007holier}, or,  negative self-esteem \cite{ellemers1997sticking}. 

Role exit theory by \kedit{Ebaugh} 
% Ebanaugh 
\cite{ebaugh1988becoming} focuses on the process of disengaging from a social role that is central to one’s identity. This theory examines the stages individuals go through as they transition out of significant roles (e.g., employee, or spouse). Works on role exit theory describe the importance of initial doubts and questioning of beliefs in triggering the role exit process \cite{dziewanski2020leaving,ebaugh1988becoming} and how the possibility of new roles or better external alternatives can facilitate the exit.   

Another approach to understanding disengagement as a socially\kedit{-}mediated process is to consider the impact of stigma \cite{goffman2009stigma} \cite{becker1963outsiders}. Stigma refers to how individuals are discredited due to attributes or behaviors that are deemed socially undesirable and can be assigned to individuals in a group by other group members. While the concept of stigma can be attached to whole groups, or even the act of leaving the group, 
% from the individual's point of view, 
exiting can be a way of managing 
% their 
\kedit{one's}
stigmatized identity \cite{gronning2013fatness,mahajan2008stigma}. 

\subsubsection{\textbf{Disaffiliation in religious groups}}
\label{lit:religion}
The process of exiting religious or other high-cost groups is commonly described as ``disaffiliation'' \cite{sandomirsky1990processes}. Theoretical scholars have contributed extensive works studying motivations and consequences of disaffiliation from religious groups across the last several decades.   

The motivations behind religious disaffiliation have been the subject of various studies revealing a complex interplay of social, psychological, and ideological factors. Individuals often leave religious affiliations to seek more physical and ideological autonomy, rejecting their community's philosophical and lifestyle restrictions \cite{beider2022motivations,coates2013disaffiliation,davidman2007characters,engelman2020leaving}. The need for philosophical autonomy may stem from the exposure to science and liberalism \cite{miles2023reasons,zuckerman2016nonreligious}, \kedit{or} from disillusionment with religious authority, particularly in more hierarchical religious organizations \cite{woodhead2016rise}.

\subsubsection{\textbf{Disengagement in extremism literature}}
\label{lit:extremism}
Disengagement from violent extremism is generally described through the push pull factors \cite{cherney2021push}.  Push factors are negative experiences or disillusionments that drive individuals away from extremist groups, while pull factors are positive alternatives that attract individuals toward a non-violent lifestyle. Peer dynamics, along with the waning trust and respect for leaders\kedit{,} 
% is one of the 
\kedit{are a} primary reason
% s 
for deradicalization in violent right-wing extremists \cite{aho1988out,ferguson2015leaving,bjorgo2009leaving,windisch2016disengagement,morrison2021systematic}. Similarly, change in personal circumstances that lead to identity transformation (e.g, shifting priorities as a new parent) also pull extremists away from their movements \cite{bjorgo2009leaving,munden2023extremist,morrison2021systematic,windisch2016disengagement}. Scholars also note the role of counter-exposure where assimilating in racially diverse communities or work environments led to deradicalization in white supremacist groups \cite{liguori2022walking}.

\subsubsection{\textbf{Organizational exit}}
\label{lit:orgs}
Organizational exit \cite{hom2011organizational}, often referred to as turnover, is the process through which employees leave an organization, either voluntarily or involuntarily. Studies looking at employee turnover concede that feeling undervalued or dissatisfied within an organization leads to negative self-esteem and increases the turnover intention \cite{dwiyanti2019job,sharma2024effect,geurts1998burnout,farrell1992exploring,hirschman1970exit}. Exhaustion and burnout caused by work related stress also influence employees' decision to exit \cite{steffens2018unfolding,mobley1977intermediate,zhou2015job,geurts1998burnout}. Presence of better alternative opportunities can further facilitate organizational exit \cite{farrell1992exploring,hom2019employee}.

\subsubsection{\textbf{Disengagement from problematic subcultures}}
\label{lit:emp}
While there is limited literature studying exit or disengagement from problematic subcultures such as conspiracy theory groups or \kedit{the} manosphere, a few qualitative works outline various factors in disengagement. Interview studies with former conspiracy theory believers find that exposure to contradictory evidence \cite{engel2023learning,xiao2021sensemaking} and getting distance from the internet and social media \cite{engel2023learning} led people out of the conspiracy theory rabbit holes. Studies analyzing discussions in \texttt{r/exredpill} and \texttt{r/IncelExit} communities revealed data that the decision to leave stemmed from the members realizing the harmful effects of red pill ideology on themselves and the people around them \cite{thorburn2023exiting}.

\subsection{CSCW and HCI perspectives on disengagement and online communities}
CSCW and HCI literature studying broader scopes of exit processes has focused largely on online recovery communities. Specifically, researchers have extensively examined how online recovery communities function as critical spaces for navigating social disengagement, identity transitions, and mental and physical health aspects of recovery. Online platforms ranging from Reddit to health forums serve as sites of both refuge and transformation, particularly for individuals distancing themselves from stigmatized identities or addictive behaviors. For instance, studies of Reddit communities like r/StopDrinking and r/OpiatesRecovery highlight how users collectively engage in meaning-making and accountability as they navigate sobriety, with implications for peer support design and integration with formal care systems {\cite{gauthier2022will}}. In parallel, work on health-related forums shows how user posts can reflect identifiable phases of disengagement or relapse, enabling the design of systems that scaffold recovery through predictive support {\cite{maclean2015forum77}}. Moreover, online self-disclosure is deeply shaped by concerns around stigma, privacy, and audience. Andalibi and Forte {\cite{andalibi2018not}} show how individuals managing stigmatized experiences—such as pregnancy loss—strategically engage with visibility and support on Facebook. 

While CSCW has extensively studied self-disclosures {\cite{de2014mental}}, {\cite{ammari2019self}}, {\cite{engel2023learning}}, collective sensemaking {\cite{xiao2021sensemaking}}, and social support {\cite{mccoy2024you}},{\cite{gauthier2022will}},{\cite{maclean2015forum77}}, less attention has been paid to the phenomenon of community exit, especially from ideological groups where identity, belonging, and moral alignment are deeply entangled. This paper builds on calls in HCI to center lived experience, particularly around values and harm, by analyzing how people narrate their exits from high-commitment belief-based communities.

%\vspace{-5pt}
\subsection{Research on extracting reasons from social media text}

RQ2 seeks to identify and categorize the reasons individuals cite for disengaging from ideological or belief-based communities, using self-narrated exit stories. This task requires surfacing underlying motivations, reasons and turning points expressed in natural language often in an informal way. 

Traditional NLP methods for extracting reasons or causal motivations from social media text relied heavily on rule-based approaches and classical supervised models such as SVMs and logistic regressions {\cite{son2018causal}}. These approaches typically used surface features, lexicons, and hand-engineered rules to identify explicit causal markers (e.g., ''because,'' ''due to'') {\cite{doan2019extracting}}{\cite{son2018causal}}. 

More recently, large language models (LLMs) such as GPT-3.5 and GPT-4 have been shown to generalize well in zero- or few-shot contexts, especially with careful prompt design {\cite{pietron2024efficient}} and multi-class setting {\cite{chen2024exploring}}.  The method used in RQ2 draws inspiration from this line of research: a zero-shot multi-label classification strategy using GPT-4 is employed to detect and score multiple disengagement factors in exit stories. By prompting GPT-4 with a theoretically grounded exit reasons framework and asking for probabilistic assessments, this approach enables interpretable, scalable labeling of both explicit and implicit motivations for exit—capturing the complexity of disengagement narratives beyond what traditional models can offer.

\section{RQ1: Identifying and charactering the types of \textit{exit stories} on Reddit}

% \usepackage{booktabs}
% \usepackage{graphicx}
\begin{table}[]
\centering
\resizebox{\textwidth}{!}{%
\begin{tabular}{@{}cccccccc@{}}
\toprule
\textbf{\begin{tabular}[c]{@{}c@{}}Exit story \\ theme\end{tabular}} & \textbf{\begin{tabular}[c]{@{}c@{}}\# \kedit{e}xit \\ subreddits\end{tabular}} & \textbf{\begin{tabular}[c]{@{}c@{}}\# total \\ posts\end{tabular}} & \textbf{\begin{tabular}[c]{@{}c@{}}\# Q\&A pairs \\ in exit subs\end{tabular}} & \textbf{\begin{tabular}[c]{@{}c@{}}\# Q\&A pairs \\ in rest of Reddit\end{tabular}} & \textbf{\begin{tabular}[c]{@{}c@{}}\# exit stories \\ BERT or Q\&A\end{tabular}} & \textbf{\begin{tabular}[c]{@{}c@{}}\# exit stories \\ sampled for analysis\end{tabular}} & \textbf{\begin{tabular}[c]{@{}c@{}}\# unique\\ authors\end{tabular}} \\ \midrule
Religio\kedit{n} & 114 & 1,156,613 & 437 & 5,742 & 96,731 & 5000 & 4217 \\
Manosphere & 3 & 7,310 & 119 & 2,854 & 1966 & 1966 & 1327 \\
Conspiracy theory & 3 & 443 & 63 & 1,036 & 851 & 851 & 414 \\
Politics & 9 & 7,486 & 53 & 311 & 765 & 765 & 652 \\
Lifestyle & 2 & 15,729 & 294 & 3,212 & 6205 & 6205 & 5081 \\ \bottomrule
\end{tabular}%
}
\caption{Table describing various data points relevant to \textit{exit stories} analyzed in this paper. I compile \textit{exit stories} from three different sources: post in exit subreddits, Q\&A in exit subreddits, Q\&A outside of exit subreddits. In total, the dataset contains 106K \textit{exit stories} obtained from RQ1 however, 15K \textit{exit stories} were analyzed in RQ2 by downsampling posts from Religion related subreddits. }
\label{tab:data_table}
\end{table}

In this section, I describe the process for identifying exit subreddits (Section \ref{rq1:exitsubs}) and classifying posts from the exit subreddits as \textit{exit stories} (Section \ref{rq1:bert}). I also detail methods for sourcing \textit{exit stories} through question and answer (Q\&A) pairs from both inside (Section \ref{rq1:exitqa}) and outside (Section \ref{rq1:redditqa}) of exit subreddits. All data identified in this section is labeled with of the five exit themes described in Section \ref{rq1:themes}.

\begin{figure*}[]
    \centering
    \includegraphics[width=0.99\textwidth]{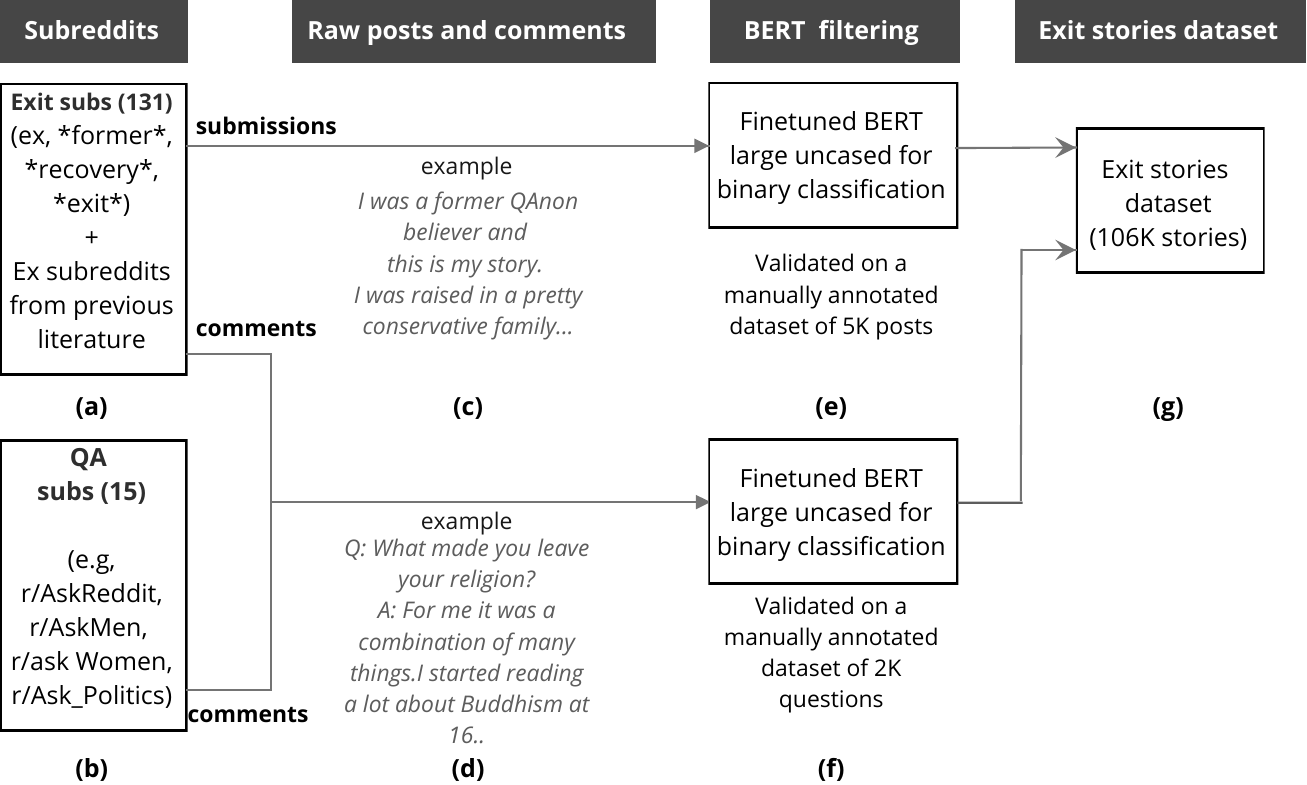}
    \caption{Figure describing the dataset and methods used in RQ1 for identifying \textit{exit stories} on Reddit. I start by first compiling a list of exit subreddits on Reddit (a) and extracting submissions from 131 identified subreddits (b). I then use a BERT-based model (e) to train and classify \textit{exit stories} in submissions from exit subreddits. Separately, I also extract Q\&A discussions in exit communities (d) and train a smaller BERT model to identify questions (or posts) asking people to share their \textit{exit stories} (f). See Figure \ref{fig:rq1_examples} for examples. This model is further used to identify similar questions on the rest of the Reddit Q\&A subreddits (b). The first order comments obtained from all Q\&A discussions are further added to the \textit{exit stories} dataset (g)}
    % \caption{RQ1 data and methods diagram }
    \label{fig:rq1_data_methods}
\end{figure*}

\label{rq1:data}
\subsection{Data}
\begin{figure*}[]
    \centering
    \includegraphics[width=0.99\textwidth]{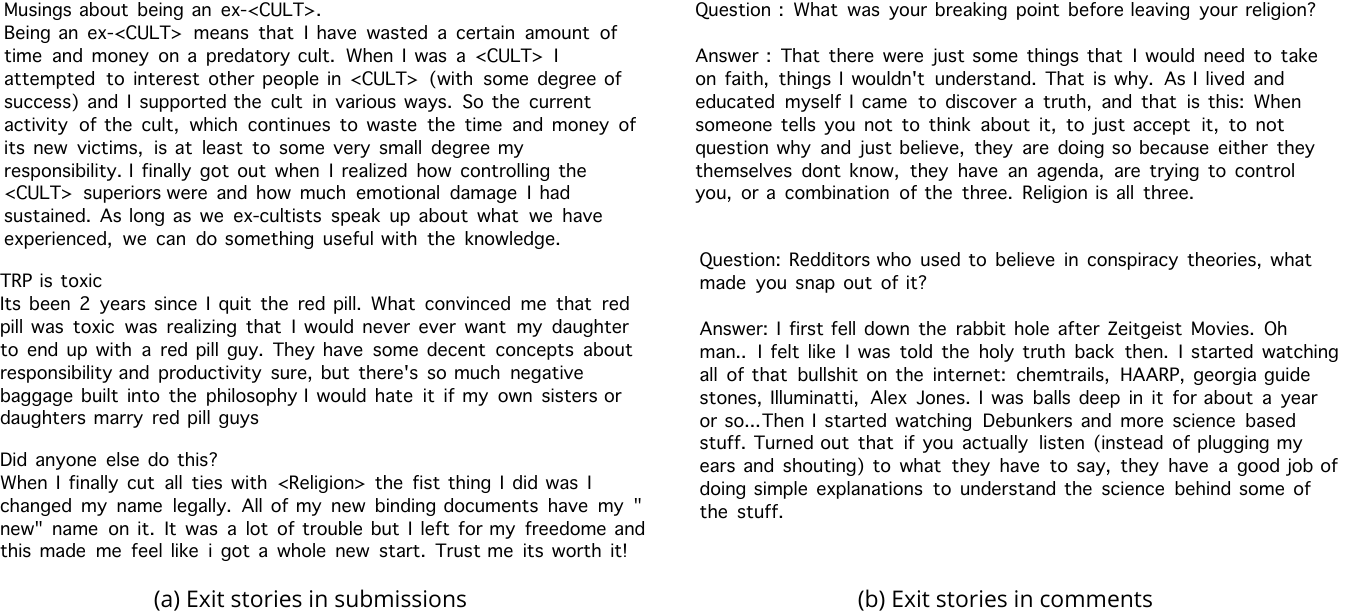}
    \caption{RQ1 examples of \textit{exit stories} on Reddit. (a) contains \textit{exit stories} obtained from submissions in the exit subreddits whereas (b) displays exit stories as answers to Q\&A both inside and outside of exit subreddits.}
    \label{fig:rq1_examples}
\end{figure*}

%\subsubsection{\textit{exit stories} on Reddit}
Reddit hosts a range of communities that support former or ``ex'' members of various social groups. For example, \texttt{r/exredpill}\footnote{\kedit{Reddit communities can be accessed by appending the subreddit to the domain. For example, \texttt{r/exredpill} is located on the web at \url{https://www.reddit.com/r/exredpill/}.}} subreddit describes itself as a \textit{forum for former redpillers and others who recognize the damage caused by redpill.}
% Or,  
\kedit{The} \texttt{r/exmormon}, 
% \footnote{\url{https://www.reddit.com/r/exmormon/}}, 
\texttt{r/exchristian}, \kedit{and}
% \footnote{\url{https://www.reddit.com/r/exchristian/}}, 
\texttt{r/exmuslim} \kedit{subreddits} 
% \footnote{\url{https://www.reddit.com/r/exmuslim/}} 
are few of the biggest communities that invite members who have disaffiliated from their religion. These communities serve as social spaces for reflection and support for those who have \kedit{or seek to} move
% d 
away from their past identities. Often, members also share \textit{exit stories} describing personal journeys out of previous affiliations. Figure \ref{fig:rq1_examples} describes some typical examples of \textit{exit stories} across various such communities. To capture the full breadth of self-disclosed exit experiences, I source exit stories from both original Reddit submissions and responses to question-and-answer (Q\&A) posts. These two formats serve different narrative and social functions: posts allow users to share unsolicited, often longer-form reflections on their exit experiences, while responses to Q\&A are typically elicited by explicit prompts, revealing how people frame their stories when invited to explain ''what made them leave.'' 
For example, in a stand-alone post in \texttt{r/exredpill}, one user writes a post titled: ``My journey through trp..''. This kind of submission, followed by a longer unsolicited disclosure, often reflects introspective storytelling. In contrast, in a Q\&A thread on \texttt{r/AskReddit} titled "Redditors who left a cult, what made you leave?", responses were shorter and more focused, shaped by the question’s framing and by the social expectation to explain a turning point. Including both posts and Q\&A increases the completeness of the exit stories dataset and provides a more holistic view of disengagement discourse, capturing both internally motivated posts and socially cued disclosures. In this section, I describe the methods used for surfacing potential \textit{exit stories} from Reddit posts and comments.  

% \subsubsection{\textbf{Potential Exit stories in Reddit submissions}}
\label{rq1:exitsubs}
\subsubsection{\textbf{Detecting subreddits with potential \textit{exit stories}}}
%\textbf{Identifying exit communities on Reddit}\\
I started by first exploring a 
% large 
list of 13 million subreddits as of December 2022 published by Pushshift \cite{baumgartner2020pushshift,pshift}. A manual analysis of some of the larger communities revealed that exit community subreddit names.
% ies 
typically start with the prefix ``ex'' (e.g, \texttt{r/exscientology}) or contain signifiers such as ``former'' (e.g, \texttt{r/FormerLeftists}), ``exit'' (e.g, \texttt{r/IncelExit}) or ``recovery'' (e.g, \texttt{r/MLMRecovery}).
% in the subreddit name.
Filtering based on 
% these 
criteria 
\kedit{of 1) subreddit names containing these terms and 2) subreddit subscriber counts containing}
% for subreddits with 
at least 100 subscribers resulted in a smaller list of 2,463 subreddits. 
I further manually inspected the ``top'' and ``hot'' feeds of each of the 2,463 subreddits to note whether the subreddit contains personal \textit{exit stories} shared by their members. I specifically focus on subreddits that record personal psycho-social journeys of disengagement and exclude subreddits meant for ex-pats (e.g, \texttt{r/exjordan} or \texttt{r/exsaudi}) that usually see more logistical discussions around the physical process of moving or exiting a location. I also further added subreddits referenced in previous works that do not follow the naming patterns described above, for example, \texttt{r/ReQovery} \cite{engel2023learning,kruglova2023understanding}. 
During analysis, each subreddit was also tagged with 
% the general category
\kedit{one of the five thematic areas} (Section \ref{rq1:themes}). For example, \texttt{r/exchristian}, \texttt{r/exmuslim} and \texttt{r/exmormon} were noted as ``Religion'' related subreddits whereas \texttt{r/exvegan}, \texttt{r/excarnivores} were tagged as ``Lifestyle'' related subreddits. Table \ref{tab:data_table} displays all categories found on Reddit along with the number of subreddits and examples of subreddits in each category. After the manual analysis described above, 131 subreddits remained in the dataset. 

After identifying exit subreddits, I collected all submissions in the subreddits 
% till 
\kedit{from} the creation of the subreddits  \kedit{through} December 2023 using the Pushshift Reddit dataset \cite{baumgartner2020pushshift} and the Reddit data dumps \cite{acaTorrents}. Table \ref{tab:data_table} displays the total number of posts collected per category from the filtered list of subreddits.

\label{rq1:exitqa}
\subsubsection{\textbf{\kedit{Detecting \textit{exit stories} from Q\&A in exit subreddits}}}
Exit subreddits surfaced in the last subsection also have a trend of question-answering where users ask questions related to peoples' exits and the Redditors share their answers in comments. I focused specifically on submissions containing Wh- questions (e.g., "Why did you leave...", "What made you change your mind...") because these types of queries are most likely to invite answers specific to the users' process of disengagement.

However, not all questions asked in exit subreddits request \textit{exit stories}. For instance, many users also make posts asking for support or resources for loved ones in recovery---``\textit{What can I do to help my sister get out of <CULT>}?''. Similarly, exit subreddits are also common places to spark philosophical debates around religion, philosophy, politics, and other relevant topics. For example, ``\textit{How do religions manage to exert so much control? In my opinion, it has something to do with our upbringing.}''

The 131 exit subreddits combinedly contained 9,324 submissions with Wh-questions. To detect all questions asking for \textit{exit stories}, I fine-tuned \texttt{BERT-large-uncased} model \cite{kenton2019bert} with a manually labeled set of 1000 submissions with Wh-questions. The training included a five-fold cross-validation strategy with 70\% - 30\% train-test split with a training batch size of 12 and across three epochs.

Each submission (or, question) was labeled as 1 or 0 based on whether it asked Reddit users to share their experiences around disengaging or exiting from the given community. 
This task involved labeling Reddit posts as questions or non-questions related to exit stories. This was a binary, syntactically driven classification based on recognizable cues such as interrogative phrasing (e.g., "why did you leave", "how did you decide to leave",  "what made you give up.."). Given the low subjectivity of the task, annotations were performed by a single coder.

Given the simplicity of the labeling task, the BERT model was able to achieve high average precision ($94.3\%$) and average recall ($89.6\%$) over a five fold cross-validation task. I used the fine-tuned model to identify \kedit{241} submissions with questions related to \textit{exit stories}. 
% Based on the model predictions, a total of 241 posts were labeled positively for exit questions. 
Out of the 241 posts, 96 had at least one comment in a reply. I extracted all first order comments---comments that are direct replies to the original submission \kedit{question}---from the 96 submissions as potential texts containing \textit{exit stories}. 

\label{rq1:redditqa}
\subsubsection{\textbf{\kedit{Detecting \textit{exit stories} from Q\&A in non-exit subreddits}}}
Reddit has a rich culture of subreddits such as \texttt{r/AskReddit} specifically designed for question-answering. I further looked outside of the 131 exit subreddits, into Reddit's biggest Q\&A subreddits to find discussions relevant to \textit{exit stories}. I started from the list of Q\&A subreddits from Reddit lists \footnote{\url{https://www.reddit.com/r/redditlists/comments/1f08tj/list_and_multireddit_of_subreddits_for_asking/}}---a crowdsourced repository of subreddits sorted into different genre\kedit{s}---and compiled a list of 15 non-topic specific Q\&A subreddits. For example, I included subreddits that allow questions for large groups of general populations such as \texttt{r/AskReddit}, \texttt{r/AskWomen}, \texttt{r/AskMen}, \texttt{r/AskAnAmerican} etc., and excluded subreddits that are focused on specific topics, such as \texttt{r/AskHistorians} or \texttt{r/AskStatistics} which are less likely to contain personal \textit{exit stories}. 

I further extracted submissions with Wh-questions from the selected subreddits and identified questions appropriate to the exit story discussions using the finetuned BERT model described in the previous subsection. This method surfaced an additional 1,474 questions with a non-zero number of first order comments. All questions were further manually verified \kedit{and annotated for relevance to the five key thematic areas of \textit{exit stories}.}

\kedit{First-order comments on these exit story-related question submissions were directly added to the \textit{exit stories} data.}

\label{rq1:themes}
\subsection{Themes in \textit{exit stories}}
The communities selected for this study were identified through a systematic sampling method aimed at capturing a diverse range of disengagement narratives. While these communities differ in their specific beliefs and cultural contexts, they share structural and functional similarities that justify their comparative analysis.

Each community operates utilizing similar affordances such as anonymity, asynchronous communication, and community moderation. These shared characteristics facilitate comparable modes of self-disclosure and narrative construction among users. Moreover, the act of disengagement across these communities involves a renegotiation of personal identity and belief systems, a process that is central to this paper.
While the paper foregrounds exits from problematic communities—such as conspiracy theory spaces, redpill forums, and ideologically extreme subcultures—it also includes stories from lifestyle based communities where the exits may be subtler or less socially stigmatized. These broader communities were retained to allow for comparative analysis of disengagement processes across both high-control and harm-associated groups and more normative lifestyle communities.

All \textit{exit stories} are labeled into one of the five themes described in this section. \textit{exit stories} obtained from the posts and comments from the 131 
% Q\&A 
\kedit{exit} 
subreddits were labeled using the theme of the subreddit. \textit{Exit stories} obtained from Q\&A on the rest of the Reddit were 
% manually 
\kedit{labeled according to comment content.}

\begin{itemize}
    \item \textbf{Religion\kedit{:}}
    % exit stories: }
    Stories by Reddit users exiting from religious or pseudo-religious groups. Notably, several of the groups involved in this category are referred to as ``cults'' by the exit story authors but are labeled as ``organizations that operate for religious and charitable purposes'' by government regulatory bodies. Examples of the subreddits in the ``Religion'' theme include \texttt{r/exmormon}, \texttt{r/exjw}, \texttt{r/exscientology} 
    \item \textbf{Manosphere\kedit{:}}
    % exit stories: } 
    \kedit{Stories by Reddit users exiting from the manosphere, a}
    % Manosphere refers to 
    loosely connected set of ideologies that promote ideas about men's rights, masculinity, and gender roles, often in reaction to feminism and societal changes that are perceived as unfavorable to men \cite{ging2019alphas}. While most of the exit stories in the dataset come from \texttt{r/exredpill} and \texttt{r/IncelExit} subreddits, the Q\&A also involves accounts from ex-MGTOW (Men Going Their Own Way) \cite{jones2020sluts} and ex-blackpill \cite{glace2021taking} members. 
    \item \textbf{Conspiracy theory\kedit{:}}
    % exit stories: } 
    \kedit{Stories by Reddit users exiting conspiratorial beliefs and communities.}
    Conspiracy theories are generally defined as theories on secret plots by powerful people or organizations influencing world events \cite{douglas2019understanding}. While Reddit has a range of subreddits dedicated to discussing various types of conspiracy theories, the exit discourse in submissions is largely limited to QAnon conspiracy theory in the \texttt{r/ReQovery} subreddit. Q\&A discussions outside of the exit subreddits also include accounts from former Flat Earth theory, pizza gate, 9/11 conspiracy theory and anti-Vaxx movement believers, 
    \item \textbf{Politic\kedit{s:}}
    % al exit stories: } 
    % Exit stories in this category come from a 
    \kedit{Stories by Reddit users exiting a }range of political beliefs including former 
    % liberals, 
    \kedit{D}emocrats, \kedit{R}epublicans, \kedit{liberals}, conservatives, and libertarians. While the exit subreddits themselves do not contain a lot of exit stories, the majority of the political exit story data comes from the Reddit Q\&A. 
    \item \textbf{Lifestyle\kedit{:}}
    % exit stories: } 
    \kedit{Stories by Reddit users exiting a particular way of living. In the data}
    there are two main lifestyle subreddits, both relevant to diets---\texttt{r/exv\kedit{e}gans} and  \texttt{r/excarnivores}. While ex-vegans contribute a majority of the exit stories in submissions, ex-carnivores voice their experiences \kedit{primarily} through 
    % Reddit 
    \kedit{comments responding to exit}
    Q\&A.
\end{itemize}

% \subsection{Identifying Exit Stories} 
\label{rq1:bert}
\subsection{\kedit{Exit stories disclosed through exit subreddit submissions}}
% Table \ref{tab:data_table} describes the number of posts and comments grouped by exit story types. 
% C
\kedit{While exit stories disclosed through Q\&A c}omments 
% obtained from Q\&A 
were directly added to the exit stories dataset (Figure \ref{fig:rq1_data_methods} (b, d, f)),
% . However, 
\kedit{those disclosed through exit subreddit} submissions 
% from exit subreddits 
(Figure \ref{fig:rq1_data_methods} (c)) needed to be classified further. 
% In this section
\kedit{Here}, I describe my process for using another \texttt{BERT-large-uncased} model fine-tuned on a manually labeled dataset of 4,656 submissions to obtain exit stories from Reddit posts. 

\subsubsection{\textbf{Manually labeling submissions in exit subreddits}}
% Along with the 
\kedit{In addition to} exit stories, submissions \kedit{within}
% from
exit subreddits span a variety of topics including support for recovery, discussions on the inner workings of various religions or social groups, and recovery updates by frequent subreddit users. To create a ground truth dataset of \kedit{non Q\&A }submissions with exit stories, I generated a stratified random sample of 4,656 posts 
% based 
from the subreddit categories \kedit{, corresponding to themes in exit stories,}---Politics, Religion, Conspiracy theories, Manosphere, and Lifestyle.
% described before. 
Each post was labeled as "Exit story" or "Non exit story". I labeled the post as an exit story if it contains
% ed 
any personal references to leaving behind an ideology, belief, social or religious group organization,  or lifestyle. Similar to the labeling of questions described earlier, this labeling task did not involve subjective judgment or interpretation. Binary, simplistic tasks such as these could be carried out by a single coder, not requiring inter annotator agreement {\cite{mcdonald2019reliability}}. Roughly 20\% of the sampled data contained exit stories. 

\subsubsection{\textbf{Building a model for classifying exit stories}}
I chose to fine-tune a BERT Large Uncased model to classify Reddit posts into exit stories. BERT is a state-of-the-art pre-trained language model that excels at understanding contexts in natural language due to its bidirectional attention mechanism \cite{kenton2019bert}. The large variant of BERT, can leverage a more complex model with 24 layers and learn deeper linguistic patterns, which is particularly beneficial for nuanced content like Reddit posts that often involve informal language and community-specific jargon. Fine-tuning a robust, low-cost and large pre-trained model like BERT strikes a balance between leveraging BERT’s pre-existing knowledge and tailoring it to the unique language patterns and vocabulary present in the exit posts.

The model was trained with a batch size of 12 and an 80-20 train-test split with 3-fold cross validation.
In 3-fold cross-validation, the model was able to achieve a satisfactory performance of average 91\% precision, 89\% recall, 94\% accuracy and 89\% F1 score indicating that the model is reliably able to predict exit stories across all subreddit categories.  
I apply the fine-tuned model on the entire dataset to compile the larger pool of exit stories. Any false positives from this model are further managed through LLM based methods described in the next section. Table \ref{tab:data_table} records the total number of exit stories either predicted with fine-tuned BERT model or obtained from the Q\&A dataset. Given the large number of contributions in religion-related subreddits, I randomly sample 5000 religious exit stories for the RQ2 analysis.

\section{RQ2: Understanding factors in disengagement in exit stories}
While RQ1 focuses on identifying exit stories, the second research question aims at generating a deeper understanding of what makes people leave their communities. To understand disengagement disclosures on Reddit, I first employ a transdisciplinary approach, integrating perspectives from sociology, social psychology, organizational behavior, economics, political science, religion, psychology, and HCI. I synthesize a content analysis framework of exit reasons---\textit{exit reasons framework}---by combining overlapping concepts from different scholarships while still preserving the unique theoretical approaches to understanding disengagement. I then refine and strengthen the framework by evaluating its applicability to exit stories compiled in RQ1. Finally, I use the framework, along with the theoretical definitions and examples generated through manual analysis, to compare and contrast various dimensions of disengaging from political, religious, manosphere, conspiracy theory, and lifestyle groups identified in RQ1. 

\subsection{Building the transdisciplinary \textit{exit reasons framework}}

% \usepackage{booktabs}
% \usepackage{graphicx}
\begin{table}[]
\centering
\resizebox{\textwidth}{!}{%
\begin{tabular}{@{}lll@{}}
\toprule
Theories or observations & Posited reasons for exit& References \\ \midrule
\begin{tabular}[c]{@{}l@{}}Social identity \end{tabular} & Group norms don't conform to the individual's identity & \cite{tajfel1978intergroup,hirsh2016mechanisms} \\ \cmidrule(l){2-3} 

 & Individuals with negative self-esteem in the group may prefer to leave & \cite{ellemers1997sticking} \\ \cmidrule(l){2-3} 
 & Individuals may leave due to moral misalignment & \cite{monin2007holier} \\ \midrule
\begin{tabular}[c]{@{}l@{}}Cost benefit models\end{tabular} & Individuals leave social groups in presence of better alternative opportunities & \cite{gabay2024social}  \\ \midrule
Cognitive dissonance & Individuals experience uncertainty or doubts due to contradictory evidence & \cite{festinger1954theory,festinger1959cognitive} \\ \midrule
Role exit theory & Exit is initially triggered by doubts and uncertainty & \cite{dziewanski2020leaving,ebaugh1988becoming}  \\ \cmidrule(l){2-3} 
 & Exit can be facilitated by better alternatives & \cite{ebaugh1988becoming} \\ \midrule
Extremism and criminology & Individuals leave organizations after moral, tactical or political disagreements& \cite{bjorgo2009leaving,morrison2021systematic}\\ \cmidrule(l){2-3}
& Individuals leave organizations after getting counter-exposure & \cite{liguori2022walking} \\ \cmidrule(l){2-3}
& Individuals leave organizations due to disappointment with peers and leaders & \cite{aho1988out,bjorgo2009leaving,windisch2016disengagement,morrison2021systematic} \\ \cmidrule(l){2-3}
& Individuals leave organizations due to change in personal circumstances and identity & \cite{bjorgo2009leaving,munden2023extremist,morrison2021systematic,windisch2016disengagement} \\ \cmidrule(l){2-3}
 & Individuals leave organizations due to excessive violence & \cite{windisch2016disengagement} \\ \midrule
Exit, Voice, and Loyalty Theory & Individuals exit organizations because of the dissatisfaction with organization & \cite{farrell1992exploring,hirschman1970exit} \\ \cmidrule(l){2-3} 
 & Individuals exit organizations for better alternative options & \cite{farrell1992exploring} \\ \midrule
Organizational exit & Individuals leave groups after emotional exhaustion and burnout & \cite{mobley1977intermediate,zhou2015job,geurts1998burnout} \\ \cmidrule(l){2-3} 
 & Individuals leave groups for better external opportunities & \cite{hom2019employee} \\ \cmidrule(l){2-3} 
 & Individuals leave groups after feeling undervalued or unsupported & \cite{dwiyanti2019job,sharma2024effect,geurts1998burnout}  \\ \midrule
% Social Network Theory & Individuals leave social networks after feeling unimportant &  \\ \cmidrule(l){2-3} 
%  & Individuals leave social networks due to weakened social ties &  \\ \midrule
% Groupthink Theory & Individuals leave due to dissenting moral or philosophical standards &  \\ \midrule
Stigma and Deviance & Individuals choose to leave after being labeled as deviant & \cite{bader2019deviance} \\ \cmidrule(l){2-3} 
 & Individuals disengage from social groups because of the stigmatized self-identity & \cite{gronning2013fatness,mahajan2008stigma} \\ \midrule
Religious disaffiliation & Individuals disaffiliate to seek more autonomy & \cite{engelman2020leaving,coates2013disaffiliation,hout2002more,vargas2012retrospective}  \\ \cmidrule(l){2-3} 
 & Individuals disaffiliate due to moral or spiritual misalignment & \cite{coates2013disaffiliation} \\ \cmidrule(l){2-3} 
 & Individuals disaffiliate due to lack of social support & \cite{coates2013disaffiliation} \\ \cmidrule(l){2-3} 
 & Individuals disaffiliate due to negative mental health & \cite{brooks2020disenchanted,miles2023reasons} \\ \cmidrule(l){2-3} 
 & Individuals leave religions due to coercive or strict control& \cite{altemeyer2010amazing,jacobs1984economy} \\ \cmidrule(l){2-3}
 & Individuals disaffiliate due to value-behavior consistency in religious doctrine & \cite{beider2022motivations,matthews2017perceptions} \\ \midrule
Social computing research & Disbelief in conspiracy theories after exposure to contradictory information & \cite{engel2023learning,phadke2021characterizing,xiao2021sensemaking}  \\ \cmidrule(l){2-3} 
 & Individuals disengage from conspiracy theories due to emotional exhaustion & \cite{engel2023learning} \\ \cmidrule(l){2-3} 
 & Individuals disengage from conspiracy theory discussions due to toxic social interactions & \cite{engel2023learning} \\ \bottomrule
\end{tabular}%
}
\caption{Table displaying various theoretical perspectives considered while designing \textit{exit reasons framework}. The theoretical integration process performed in RQ2 involves synthesizing exit reasons mentioned across various theories to build a consolidated content analysis framework for analyzing \textit{exit stories}}
\label{tab:exit_theories}
\end{table}

\subsubsection{\textbf{The benefits of transdisciplinary approach}}
Disengagement as a socio-psychological process is often studied through distinct disciplinary lenses. For example, social psychology examines disengagement through the lens of group dynamics and social identity theory \cite{tajfel1978intergroup,hirsh2016mechanisms}, focusing on how the group affects an individual's sense of belonging. Economics and organizational behavior theories, on the other hand, provide an individual-focused perspective and describe how people may leave organizations through a rational cost-benefit analysis \cite{becker1976economic} based on their satisfaction or dissatisfaction with the group \cite{hirschman1970exit}. While these disciplinary perspectives offer valuable theoretical tools, the lived experiences of disengaging from different types of groups can be complicated. Consider the following exit story from a Reddit user who left behind their red pill beliefs:

\begin{quote}
    \textit{Red pill almost destroyed me. \\ After weighing pros and cons of red pill I decided that the negatives outweigh the positives. First, all of these guru people are single and bitter and I started realizing how quickly they would spit out the pill if they ever experienced healthy romantic attention.
    Talking to people on trp made me feel isolated where I became more lonely (like really lonely) and miserable. I spent hours watching this shit and still ended up feeling bad for not being able to relate to the toxicity. I just want to move on in to a healthy way with all lessons learned.}
\end{quote}

This narrative can be looked at through various theoretical lenses. The author's statement about feeling lonely and not being able to relate to the red pill community suggests identity conflict as one of the reasons for the author's exit \cite{tajfel1978intergroup}. The author also mentions weighing pros and cons of the red pill ideology, signaling a rational cost-benefit analysis \cite{becker1976economic} of exiting the red pill community. Moreover, the author calls to attention the perceived hypocrisy \cite{festinger1959cognitive} of the red pill gurus while describing the negative aspects that motivated their exit. This story of disengagement exemplifies how complex social exit stories transcend single disciplinary perspectives. The author’s decision to leave is not purely a rational cost-benefit calculation,  or not purely driven by emotional burnout. Rather, it involves a combination of identity conflict, emotional exhaustion, social isolation, and ideological disillusionment. To provide a deeper understanding of such accounts recorded in the exit stories, I utilize a transdisciplinary framework which is described in the next section. 

\subsubsection{\textbf{Creating and refining the \textit{exit reasons framework}}}
Table \ref{tab:exit_theories} records various theoretical perspectives that discuss the process of individuals disengaging from groups (See the Background section for a detailed description). I follow a structured process for first creating and then refining the transdisciplinary framework for exit reasons. I first conduct a thematic review of various theories or concepts that discuss the process of disengaging from social groups, organizations, social networks, beliefs, or ideologies. I investigate specific works studying the general concept of disengagement or exit associated with various broader theoretical paradigms. Several themes span across various disciplinary perspectives. For example, frameworks from social psychology, organizational behavior, and religion all describe negative mental health or exhaustion as one of the reasons for exit. Similarly, studies in extremism and criminology, role exit, as well as cognitive dissonance, mention doubts and counter-exposure as triggers for disengagement. 

I first started by combining overlapping themes from different theoretical paradigms to generate the initial \textit{exit reasons framework}. The initial framework had 15 categories derived from combining similar exit reasons displayed in Table \ref{tab:exit_theories}. For example, concepts of better alternative/external options from a cost-benefit perspective \cite{gabay2024social}, role exit theory \cite{ebaugh1988becoming}, exit voice loyalty theory \cite{farrell1992exploring}, and organizational exit \cite{hom2019employee} were combined into one ``better external alternatives'' category. I then used the initial framework to analyze a stratified sample of 250 exit stories to assess its applicability and coverage. During the annotation, I iteratively used deductive and inductive reasoning processes \cite{fife2024deductive} to further collapse or divide categories based on the data. For example, ``doubts and uncertainty'' and ``counter exposure'' were initially two different categories that were combined into a single category of ``Exposure to contradictions'' because of their frequent co-occurrences in the data. See below for examples:
\begin{quote}
    \textit{I started having doubts about the doctrine when I first noticed how different members of the governing body had different explanations about the commandments}
\end{quote}
or,
\begin{quote}
    \textit{..over the time I became more uncertain about my faith. I was being told that all apostates were eternally doomed but some of my cousins on the outside who left seemed really happy. }
\end{quote}

On the other hand, the initial category of ``identity conflict'' was split into two categories---``identity conflict'' and ``moral or philosophical conflict''--as the exit story authors seemed to make clear distinctions between the two. For example:

\begin{quote}
    \textit{Did anyone of you find out that you were actually repressing same-sex attraction while in trp? I never realised this during my teens, maybe because I made a barrier of some kind but my relationships with girls suffered from this. When I finally started identifying myself as gay, I also started coming out of the red pill. This was all supported through the rather liberal and very LGBT-friendly environment at my uni. }
\end{quote}

and,

\begin{quote}
    \textit{Ex piller here. I swallowed the pill so hard that I started hating everyone who was supposed to be ``weak''. I had a girlfriend and I treated her like shit. I broke up with her and said all kinds of disgusting things just because she had a couple of close gay friends. But after the breakup, it all just started feeling wrong, and I am only now realizing after coming out how evil trp had made me. }    
\end{quote}

The codebook refining process also focused on disambiguating finer categories that may appear conceptually related but serve distinct analytical purposes. For instance, while ''Calls for violence'' and ''Moral or philosophical conflict'' can co-occur—since violence is often viewed as morally objectionable—they are treated as separate categories to reflect different disengagement triggers emphasized in user narratives. Calls for violence captures instances where disengagement is explicitly prompted by the group’s endorsement or incitement of violence or abuse, often marking a clear threshold moment for the author. In contrast, Moral or philosophical conflict includes broader value-based disagreements, such as discomfort with misogyny, homophobia, racism, or perceived hypocrisy, which may not involve physical violence. Separating these categories enables more precise analysis of the specific factors that users identify as motivating their exit.

During the first annotation phase, I also looked for newer categories based on the data that were not directly represented in Table \ref{tab:exit_theories}. I added one new category of ``Negative physical health'' observed in ex-vegan and ex-carnivore exit stories. The first round of annotation resulted in an \textit{exit reasons framework} consisting of 12 categories. I further ensured the stability of the framework by annotating another stratified sample of 250 exit stories. The second annotation phase did not result in any new changes to the framework. 

While the initial version of the framework was grounded in various theoretical perspectives, the exit reasons framework was not developed in isolation from data. The theoretical overview provided a coherent conceptual base point for identifying disengagement mechanisms relevant to different types of communities which was later enhanced by incorporating cues from data across two annotation rounds. This apporach is consistent with a typical theory grounded content analysis codebook development process through inductive-deductive iterations \cite{phadke2020many,decuir2011developing,oliveira2023developing}.
The final \textit{exit reasons framework} is presented in Table \ref{tab:reasons_definitions}. In the next section, I outline methods used to apply the \textit{exit reasons framework} on the exit stories dataset using GPT-4. 

\vspace{5pt}
\noindent \textbf{\textit{Notes on labeling of exit posts: }}For RQ2, a stratified sample of 250 Reddit posts was annotated with one or more exit reasons using the exit reasons codebook described above. For each annotation task, I considered whether each of the codes from the exit reasons framework, as defined in Table {\ref{tab:reasons_definitions}} is present in a given text as a reason for the author's exit. The annotation was performed by a single coder with deep familiarity with the theoretical framework and prior literature informing the labels. This approach emphasizes consistency and interpretive alignment over inter-annotator agreement. In CSCW and HCI, the use of a single coder is well-established for qualitative coding tasks requiring theoretical sensitivity or subject-matter expertise {\cite{mcdonald2019reliability}}. However, single-coder annotations may introduce biases which are acknowledged in the limitation section.

% \usepackage{booktabs}
% \usepackage{graphicx}
\begin{table}[]
\centering
\resizebox{\textwidth}{!}{%
\begin{tabular}{@{}lll@{}}
\toprule
\textbf{Exit reasons} & \textbf{Definition} & \textbf{Short form} \\ \midrule
Identity conflict & \begin{tabular}[c]{@{}l@{}}The individual’s personal or social identity is at odds with the identity promoted by the group.\\ This conflict may arise from a misalignment between the group’s norms or behaviors and the \\ individual’s existing or evolving sense of self.\end{tabular} & IDENTITY \\ \midrule
\begin{tabular}[c]{@{}l@{}}Moral or \\ philosophical conflict\end{tabular} & \begin{tabular}[c]{@{}l@{}}Individual’s ethical, moral or philosophical beliefs clash with the group’s principles \\ or commonly accepted beliefs.\end{tabular} & MORAL \\ \midrule
Negative self-esteem & \begin{tabular}[c]{@{}l@{}}Belonging to a group may lead to feelings of inadequacy, failure, or worthlessness. \\ The group environment may diminish an individual's self-esteem, through criticism, \\ comparison, or unrealistic expectations.\end{tabular} & NEG\_SELF \\ \midrule
Negative mental health & \begin{tabular}[c]{@{}l@{}}Belonging to a group results in increased stress, anxiety, depression, exhaustion, \\ burnout or other mental health challenges.\end{tabular} & NEG\_MENTAL \\ \midrule
Negative social interactions & Group members display exclusionary, undesirable, unpleasant or toxic towards an individual. & NEG\_SOCIAL \\ \midrule
Negative group image & Group’s image deteriorates due to unethical behavior, controversy, or association with undesirable traits. & NEG\_GROUP \\ \midrule
Negative physical health & Belonging to a group results in  physical exhaustion, illness, or declining health. & NEG\_PHYSICAL \\ \midrule
Calls for violence & Group advocated for violent or abusive behavior. & VIOLENCE \\ \midrule
Better external options & Individual perceives better opportunities, communities, or support systems outside the group. & EXTERNAL \\ \midrule
Coercive or strict control & \begin{tabular}[c]{@{}l@{}}Groups exert excessive control over the individual’s actions, thoughts, or relationships \\ creating an environment of coercion or oppression.  Individuals may leave to pursue freedom.\end{tabular} & CONTROL \\ \midrule
Contradictory exposure & \begin{tabular}[c]{@{}l@{}}The individual is exposed to new ideas, experiences, information, evidence, \\ or worldviews that contradict the group’s core beliefs. This may result in doubt or uncertainty.\end{tabular} & EXPOSURE \\ \midrule
Personal or world events & \begin{tabular}[c]{@{}l@{}}The individual goes through significant personal life changes or global events \\ altering priorities, ability or desire to participate.\end{tabular} & WORLD \\ \bottomrule
\end{tabular}%
}
\caption{\textit{Exit reasons framework} with 12 categories derived from theoretical integration (Table \ref{tab:exit_theories}) and iterative coding on the sample of \textit{exit stories}.}
\label{tab:reasons_definitions}
\end{table}

\subsection{Analyzing exit stories through \textit{exit reasons framework}}
\subsubsection{\textbf{GPT-4 as a zero-shot multi label classifier}}
To understand factors in disengagement from exit stories at scale, I apply the \textit{exit reasons framework} (Table \ref{tab:reasons_definitions}) in a multi-label classification task. Meaning, each exit story is classified into one or more exit reasons categories displayed in Table \ref{tab:reasons_definitions}. 

Given the highly variable linguistics styles found in exit stories, I use the OpenAI GPT-4o model which is specifically suited for high-level linguistic inference tasks. For instance, GPT-4 has shown promising performance in complex tasks such as destigmatizating online conversations \cite{bouzoubaa2024words}, analyzing online self-disclosures \cite{bouzoubaa2024decoding}, detecting and discriminating across various problematic speech categories \cite{li2024hot}. While choosing between zero-shot or few-shot prompting I considered several factors. First, given the multi-label (12 class) classification task, providing representative examples in a few shots is challenging. The role of examples provided in the few shots is to provide a good representation of, and discrimination between, various classes. However, in multi-label classification, a good representation would also mean providing examples that capture both, the uniqueness of each class along with the co-occurrences of various classes, which amounts to a large number of possible combinations and associated examples. Second, few-shot learning often struggles with generalization when the number of classes increases because a few examples might not capture the full variability of data within each class \cite{hou2021few}. Examples provided may also create confusion between labels that co-occur together \cite{thaminkaew2024prompt} and significantly increase the number of input tokens required to label a single post. Given these considerations, and the promising ability of zero-shot learning to generalize over complex data \cite{kojima2022large,wei2021finetuned}, I opted for zero-shot multi-label classification.

\subsubsection{\textbf{Selecting probability thresholds}} Given the complex nature of the task, I add in more explainability in GPT-4 labels by asking the model to provide probability scores between 0 to 1 for each label \cite{li2024hot}. 

Considering that GPT-4 performs binary classification for each provided label, it is important to determine the appropriate probability threshold for the \emph{presence} of the category that best aligns with the human judgment. Following the procedure outlined by Li \cite{li2024hot}, I observe variability in classifier performance (macro F1 scores) produced by changing probability thresholds in the range of 0.5 to 0.90 and select probability thresholds that produced the highest macro F1 scores for each category with the manually annotated data of 250 exit stories. Probability thresholds for all categories lie between 0.6 to 0.8 with most centering around 0.7. I further test the generalizability of selected thresholds by calculating macro F1 scores on the remaining 250 annotated samples and find consistent performance. See Table\ref{tab:probability_threasholds} for details on the performance across various probability thresholds and samples. Note that any false positives accumulated through the classification model can be further removed through probabilistic labeling with high probability threshold. 

% \usepackage{booktabs}
% \usepackage{graphicx}
\begin{table}[]
\centering
\resizebox{\textwidth}{!}{%
\begin{tabular}{@{}llllllcc@{}}
\toprule
 & \multicolumn{5}{c}{\textbf{\begin{tabular}[c]{@{}c@{}}Sample 1 macro F1 Scores\\ Used to fine tune prob\end{tabular}}} & \textbf{\begin{tabular}[c]{@{}c@{}}Probability threasholds\\  with best scores\end{tabular}} & \textbf{\begin{tabular}[c]{@{}c@{}}Sample 2 macro F1 Scores\\ with best threashold\end{tabular}} \\ \midrule
 & \multicolumn{1}{c}{{\color[HTML]{656565} \textbf{0.5}}} & \multicolumn{1}{c}{{\color[HTML]{656565} \textbf{0.6}}} & \multicolumn{1}{c}{{\color[HTML]{656565} \textbf{0.7}}} & \multicolumn{1}{c}{{\color[HTML]{656565} \textbf{0.8}}} & \multicolumn{1}{c}{{\color[HTML]{656565} \textbf{0.9}}} &  &  \\ \cmidrule(l){2-8} 
IDENTITY & 0.73 & 0.75 & \textbf{0.81} & 0.78 & 0.62 & 0.7 & 0.79 \\
MORAL & 0.84 & 0.86 & \textbf{0.87} & 0.87 & 0.52 & 0.7 & 0.85 \\
NEG\_SELF & 0.79 & 0.81 & \textbf{0.84} & 0.83 & 0.70 & 0.7 & 0.80 \\
NEG\_MENTAL & 0.81 & 0.87 & \textbf{0.88} & 0.79 & 0.53 & 0.7 & 0.89 \\
NEG\_SOCIAL & 0.70 & \textbf{0.75} & 0.73 & 0.65 & 0.58 & 0.6 & 0.74 \\
NEG\_GROUP & 0.55 & 0.79 & \textbf{0.83} & 0.64 & 0.58 & 0.7 & 0.80 \\
NEG\_PHYSICAL & 0.83 & 0.86 & 0.87 & \textbf{0.90} & 0.50 & 0.8 & 0.92 \\
VIOLENCE & 0.62 & \textbf{0.76} & 0.75 & 0.70 & 0.67 & 0.6 & 0.70 \\
EXTERNAL & 0.47 & 0.55 & 0.61 & \textbf{0.62} & 0.50 & 0.8 & 0.63 \\
CONTROL & 0.76 & 0.81 & \textbf{0.83} & 0.78 & 0.51 & 0.7 & 0.79 \\
EXPOSURE & 0.61 & 0.70 & \textbf{0.76} & 0.71 & 0.67 & 0.7 & 0.78 \\
WORLD & 0.62 & 0.67 & \textbf{0.70} & 0.51 & 0.49 & 0.7 & 0.65 \\ \bottomrule
\end{tabular}%
}
\caption{Table capturing macro F1 score performance for different categories across different probability thresholds. First, I calculate macro F1 scores by comparing GPT-4 generated probabilities with manually annotated sample of 250 exit stories. Next I select the probability thresholds that provides best F1 score for each category and then calculate new macro F1 scores for a held out sample of another 250 posts. Macro F1 scores from sample 1 and sample 2 are comparable to each other.}
\label{tab:probability_threasholds}
\end{table}

\begin{figure*}
    \centering
    \includegraphics[width=0.9\textwidth]{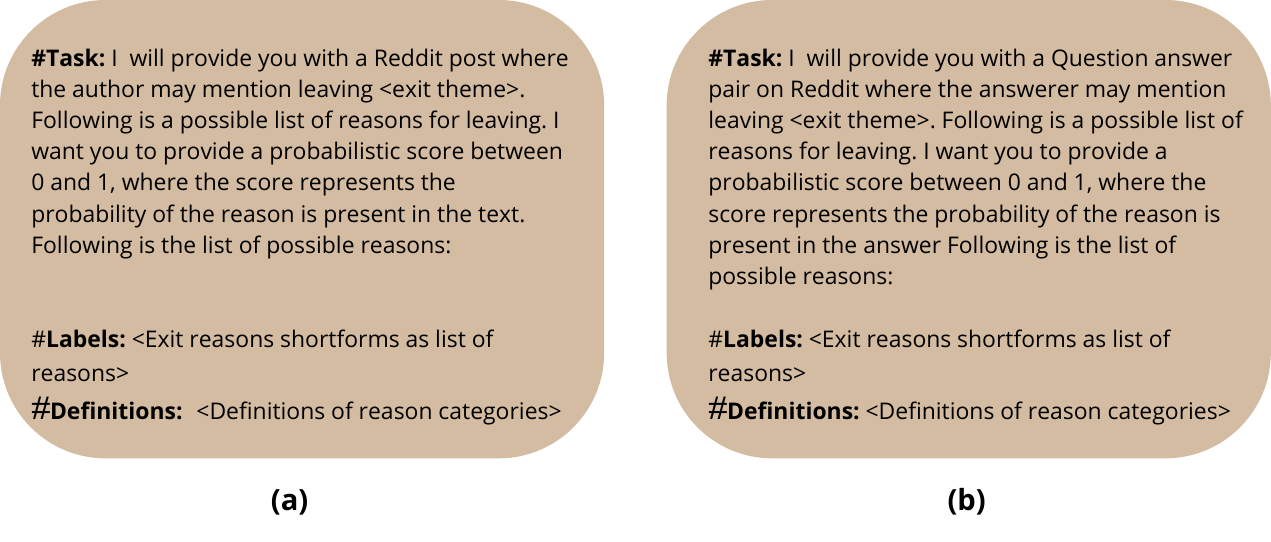}
    \caption{Figure displaying two different prompts used to label (a) \textit{exit stories} from Reddit posts and (b) \textit{exit stories} from Reddit Q\&A}
    \label{fig:prompts}
\end{figure*}

\subsubsection{\textbf{Setting up prompts}}
The prompts shown in Figure {\ref{fig:prompts}} (a) and (b) were designed for use with the OpenAI GPT-4 API. These prompts were programmatically provided to GPT-4 to classify exit stories into one or more categories from the exit reasons framework. Figure {\ref{fig:prompts}} (a) was used for Reddit submission posts, and Figure {\ref{fig:prompts}} (b) was used for question-answer pairs. Each prompt included the list of labels and their definitions and was tailored to the thematic context (e.g., conspiracy, religion, manosphere) of the story to improve contextual understanding by the model.

\subsection{Results: Disengagement factors in exit stories}

\begin{figure*}
    \centering
    \includegraphics[width=\textwidth]{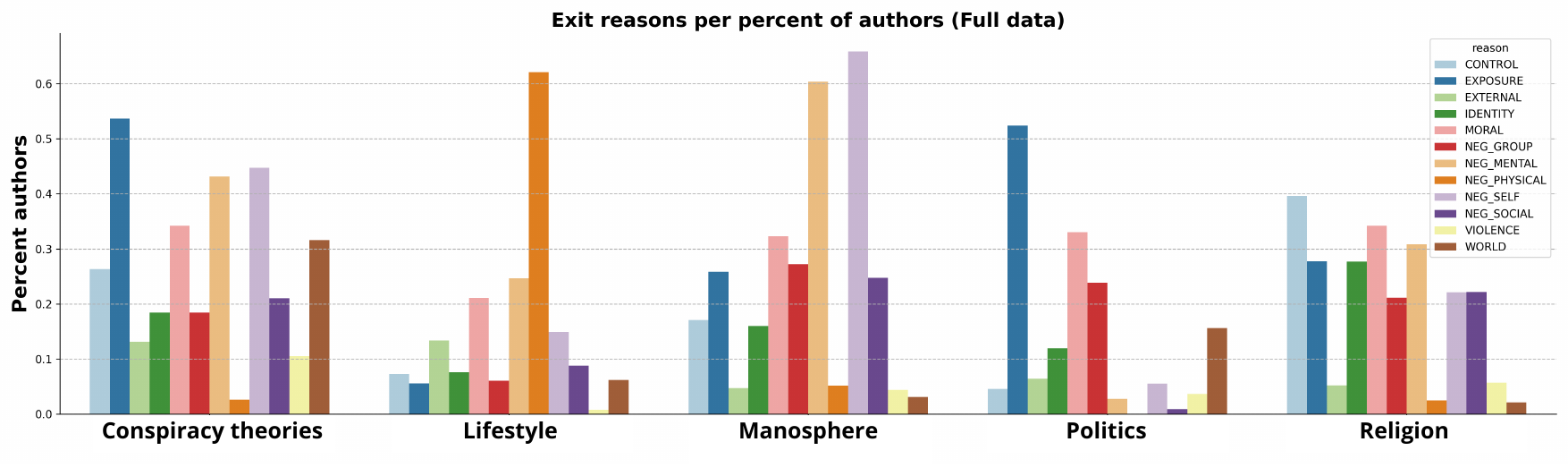}
    \caption{Figure displaying the prominence of various factors from the \textit{exit reasons framework} across various themes in the \textit{exit stories}. Overall, exits from conspiracy theory beliefs are frequently driven by exposure to contradictory information, declining mental health, and moral conflict. In contrast, exits from the manosphere communities are more often influenced by negative self-evaluation. While political ideology changes are primarily driven by counter-exposure, exits from religion are often caused by the desire for freedom.}
    \label{fig:reasons_fig}
\end{figure*}

\begin{figure*}
    \centering
    \includegraphics[width=0.9\textwidth]{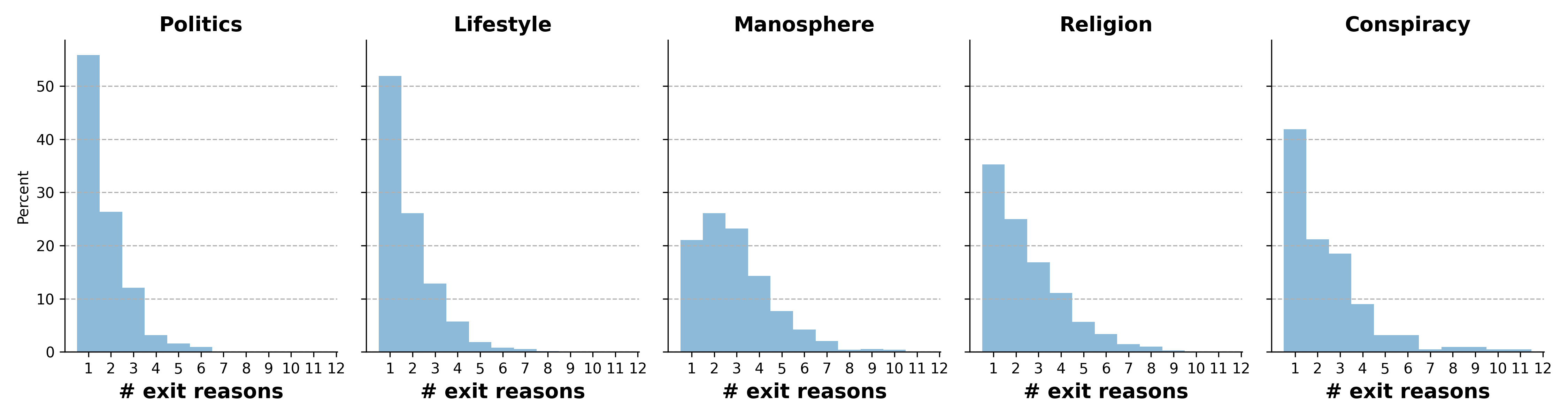}
    \caption{Figure displaying number of reasons mentioned together in \textit{exit stories}. in religious, manosphere and conspiracy theory \textit{exit stories}, more than 60\% of authors specify more than one reasons for their exit. In contrast, more than 50\% of exit authors were singularly driven by either contradictory evidence or negative physical health in political and lifestyle \textit{exit stories}.}
    \label{fig:number_reasons}
\end{figure*}

Figure \ref{fig:reasons_fig} displays an overview of how various factors of disengagement from the \textit{exit reasons framework} appear across different types of \textit{exit stories}. In general, distinctive disengagement factors are visible across various themes of \textit{exit stories}. Figure \ref{fig:number_reasons} also describes the distribution of the percentage of authors and number of exit reasons provided in their stories. Overall, at least 60\% of the stories from religion, conspiracy theories and the manosphere contain more than one exit reasons. Below, I present the results for each of the exit story themes in detail. Specifically, I present insights based on the manual analysis of a sample of 10\% of \textit{exit stories} belonging to each combination of the subreddit type and exit reason discussed.

\subsubsection{Factors in exit from conspiracy theory beliefs}
Close to 60\% of the conspiracy theory \textit{exit stories} include more than one factor from the \textit{exit reasons framework}s as reasons for disengagement. 
Nearly $53\%$ of the conspiracy theory exit story authors disclose ``contradictory exposure'' as one of the reasons for disbelieving in conspiracy theories. Many of the accounts mention getting exposed to contradictory evidence or information by either self-driven research or an intervention by close friends or family. 

\begin{quote}
    \textit{...I started doing some research and found out it's all nonsensical.} 
\end{quote}

\begin{quote}
    \textit{...I got a new girlfriend around that time and I started saying all the pizzagate, NSA leaks bs to her. I was shocked how good she was at debunking it all so easily and then I finally started seeing that I had been duped.}
\end{quote}

Similarly, more than $40\%$ authors mention mental health and negative self-image as drivers for disengagement. Upon further qualitative analysis of the \textit{exit stories}, I noticed that negative mental health usually resulted in authors wanting distance from the conspiracy theory content.

\begin{quote}
    \textit{in my last days, I was invited to a facebook group and would spend too much time on it, for probably about a solid month. and it sucked the life out of me reading all of these disgusting allegations. I hid the group to give myself a break..}
\end{quote}

Whereas feelings of shame or embarrassment, along with the realization of antisocial traits developing in themselves, usually led to negative self-esteem in authors prior to disengagement.

\begin{quote}
    \textit{...the final straw was realizing how I had started acting like a ``Karen'' around the time the first vaccines were rolled out}
\end{quote}

Notably, more than $30\%$ of the authors also mention world events such as COVID-19 pandemic and recent socio-political events as triggers for changing their worldviews. 
These findings around contradictory evidence, mental health consequences, and the role of global events all resonate with previous interview studies investigating points of fracture in conspiracy theory beliefs \cite{engel2023learning,xiao2021sensemaking}.  

As noted in the disengagement and deradicalization literature \cite{bjorgo2009leaving,morrison2021systematic}, nearly $30\%$ of the authors were driven away from conspiracy theories due to moral or philosophical conflict. 

\begin{quote}
    \textit{..I could only take so much of the constant vitriol, dehumanizing attitudes, fear-mongering, and bloodlust for people the <CULT> didn't agree with..}
\end{quote}
Interstingly, more than $20\%$ of the authors also describe their conspiracy theory beliefs as compulsive or controlling. Realizing the power they had given away to these beliefs also encouraged some authors to get out of the rabbit hole. One author mentions perceiving coercive tactics used inside Qanon as similar to ``Jonestown cult'' \footnote{\url{https://en.wikipedia.org/wiki/Jonestown}}. Another author claims feeling as if their mind and thoughts were being controlled by the excessive fear conspiracy theories instilled in them.  

Overall, a combination of mental health, self-image, moral standards, and cross-cutting exposure appear as prominent factors in disengaging from conspiracy theory beliefs. I discuss the relevance of these results in designing effective interventions in the Discussion section. 

\subsubsection{Factors in exit from Manosphere}
\textit{exit stories} in the manosphere category are largely contributed by authors leaving red pill and incel communities. Negative self-esteem ($63\%$) and negative mental health consequences ($58\%$) are two of the most prominent factors in the manosphere disengagement stories. Many users mention that red pill ideology and tactics initially improved their lives but eventually caused severe consequences. 

\begin{quote}
    \textit{ ..whenever I entered red pill spaces, I would constantly get in my head that I'm not masculine or manly enough in some facet of my life and I don't deserve companionship until I reach some imaginary standard of masculinity..}
\end{quote}

\begin{quote}
    \textit{My internalized anger towards myself made it easy to label myself as a ``beta'' or ``incel'' when in reality I had unresolved traumas and untreated ADHD.}
\end{quote}

Several \textit{exit stories} detail how before quitting, the authors reached rock bottom and experienced suicidal ideation or a lack of will to live due to the teachings of the red pill ideology. While the exit subreddits provide an open and welcoming place for such individuals, these results indicate that ex-manosphere communities can greatly benefit from mental health resources beyond peer support. I discuss opportunities around designing mental health support for exit communities in the Discussion section.
Close to $30\%$ of the manosphere \textit{exit stories} contain references to moral conflict faced by the authors. Specifically, the authors mention struggling with the morality of dating strategies, misogynistic outlook, and ideals of the ``alpha'' male promoted by the red pill. 

\begin{quote}
    \textit{..i found myself agreeing on many points, like self-improvement, taking care of my mental health... but i didn't agree with their misogynistic takes..}
\end{quote}

$26\%$ of the authors also mention disappointment with red pill influencers and red pill role models like Andrew Tate as reasons for leaving the movement. 

Qualitative analysis of red pill \textit{exit stories} also revealed that several authors find it difficult to completely get out of the red pill mindset. \textit{exit stories} detailing why the authors initially left behind the red pill also confess that various events post-departure cause them to fall back to the old mental patterns. This highlights the importance of exit communities even more, where users who struggle with recovery can seek support and encouragement from others. 

\subsubsection{Factors in exit from political ideologies}
Interestingly, most of the \textit{exit stories} discuss switching from one political ideology to other, rather than completely disengaging from politics. Two prominent reasons for switching ideologies are cross-cutting exposure ($48\%$) and moral conflict ($36\%$). Exposure to new ideas, information or worldviews responsible for switching political ideologies was mediated by a lack of unrealized political promises and general political commentary on the internet. 

\begin{quote}
    \textit{I'm currently a <POL PARTY>, that is about to commit to the <POL> party. I'm sick and tired of promises of my party and ``waiting'' for big corporations to find it in their heart to raise wages and ``appreciate'' their employees all while raking in huge profits every year. I realized it is never going to happen.}
\end{quote}

\begin{quote}
    \textit{ I'm actually in the process of switching from <POL PARTY> to <POL PARTY>. Learning fully and completely about taxes (tax rate, tax code, how much we are actually taxed) and the misleading language used by <POL PARTY> made me confused about my beliefs}
\end{quote}

Most of the \textit{exit stories} with moral conflict analyzed qualitatively were from former supporters discussing their disagreements with racism, pro-gun narratives, inflammatory language used by political leaders, corruption, religious extremism and human rights. Out of the 40 political \textit{exit stories} with moral conflict, I found two accounts by former Democrats, both indicating a move to the center, noting moral objections to overly politicizing humanitarian issues. 

Interestingly, $28\%$ of the accounts also discuss stigma related to different political parties as reasons for changing political ideologies. One user specifically mentions making a decision to switch their support due to the negative online rhetoric associated with party supporters. 

\begin{quote}
    \textit{tbh, I don't think i am smart enough to really understand why one party is better than the other. But everytime i go online, there is so much bad shit about republicans. This is the only reason why I think I stopped voting republican}
\end{quote}

\subsubsection{Factors in exit from religions}
Results from the religious \textit{exit stories} largely align with theoretical factors discussed in the disaffiliation literature. Coercive or strict control was one of the main factors expressed in religious disaffiliation. The authors mention experiencing various degrees of control---from physically isolated environments to spiritual or emotional control---that led them out of their religions. 

\begin{quote}
    \textit{..They acted like it was harmless and just a suggestion given by old men. Fuck that. Leaders, parents, and boys I dated, used those standards to shame and control my body.}
\end{quote}

\begin{quote}
    \textit{..afterwards I realized that they had no concept of individual freedom. I was forced to attend <RELIGIOUS SERVICE>, forced to hang out around specific people, forced to denouce what they considered as evil. I just wanted to be free of it all}
\end{quote}

Religious \textit{exit stories} authors also mention disaffiliating because of moral reasons ($36\%$) such as mistreatment and demonization of outsiders, sexism and misogyny, and the lack of transparency around misconduct. Close to $30\%$ of the authors describe how their decision to leave was motivated by the contrast between their self-identity and the norms enforced by their religious groups.

\begin{quote}
    \textit{I threw myself into my faith and got <religious act> as soon as I could so that I could keep homosexuality at bay. Spoiler alert: it didn't work. When I turned 18, I moved out and since I finally had the right to privacy, I started doing research about sexuality and religion and every other topic under the sun that the elders tell you not to google. I googled it all and I learned so much and I began to finally love myself for who I was}
\end{quote}

\subsubsection{Factors in exiting from lifestyles}
Most of the lifestyle \textit{exit stories} in the dataset come from ex-vegans and ex-carnivores. Exit from veganism or other kinds of diets is mainly motivated by poor physical health ($60\%$). 

\begin{quote}
    \textit{I have no energy, I'm anemic, my hair is falling out, I've developed neurological problems/heart problems apparently due to B12 deficiency (despite supplement) you get the point. Saw a nutritionist to find out how to keep eating vegan and they told me I cannot keep eating like this. }
\end{quote}

Both ex-vegans and ex-carnivores provide negative physical health as a primary driver in their decision and in $30\%$ of the accounts, poor physical health also co-occurs with negative mental health. Especially in ex-vegan stories, users discuss struggling to reconcile their stance around animal cruelty and environmentalism with their desire to lead a healthy life. Apart from the reasons outlined in the \textit{exit reasons framework}, ex-vegans and ex-carnivores also mention financial strain caused by their lifestyle (e.g, vegan products are expensive or ordering meat in restaurants costs more) as one of the reasons for switching their diet. Moreover, several accounts mention desire to feel more joy in life by switching diets, indicating that exit authors from this theme experience the process of disengagement a lot differently than other groups discussed above.  

\section{Discussion}

A key contribution of this work lies in identifying which factors drive disengagement across different types of communities—and how complex or multifactorial those exits are. These patterns are not just descriptive; they offer insight into the underlying dynamics of ideological exit and suggest design considerations for interventions, recovery support, and moderation practices.

Understanding which exit reasons dominate within each domain helps reveal the nature of tensions users experience within those communities. For example, conspiracy theory exits are often driven by contradictory exposure and moral discomfort, whereas manosphere exits are shaped more by negative self-image and mental health strain. These distinctions matter: they suggest that different communities operate through different logics of attachment—and disengagement—and thus require community-specific support systems. For CSCW researchers and designers, this opens up opportunities to create tailored resources that address the emotional, moral, or epistemic dimensions specific to each group.

In parallel, examining the number of exit reasons cited helps understand whether disengagement is experienced as a singular rupture or a gradual process. In groups like religion, conspiracy, and the manosphere, most users cited multiple co-occurring reasons—indicating that exits from these communities may often unfold over time, shaped by overlapping identity conflicts, emotional strain, and value misalignment. This has implications for platform design and intervention strategies: disengagement support should not be oriented solely around "conversion points" or single moments of doubt, but should instead account for the longitudinal nature of leaving, recovery, and reintegration.

\subsection{Disengagement as a multi-faceted process}

Whether from conspiracy theories, religions, or the manosphere, the results of RQ2 show that disengagement or exit is a complex multifaceted process. In conspiracy theory and manosphere \textit{exit stories}, more than 60\% of the authors mention multiple reasons for leaving their ideologies. Further manual analysis also reveals that their exit experiences are often triggered by personal circumstances or world events. The empirical results from RQ2 highlight that disengagement is often not just a momentary decision but often involves a prolonged, multi-stage process that varies based on individual experiences and community contexts. It may involve emotional \cite{hirschman1970exit,hom2011organizational}, cognitive\cite{festinger1954theory}, and social components \cite{morrison2021systematic}, as individuals reassess their identities, relationships, and beliefs. 

Looking at the prominent exit reasons in each community, we can also understand how the authors viewed the group they belonged to. Apart from the role of information, a significant amount of conspiracy theory \textit{exit stories} also highlight negative self-image and negative mental health. Similarly, manosphere \textit{exit stories} stress the negative self-image created by the red pill ideology. This indicates that members of conspiracy theory groups and manosphere look at their previous communities through a deeper emotional lens, going beyond the information environments that such communities exist in. \kedit{This echos sentiments expressed by ex-believers exiting problematic beliefs and communities with respect to the role of community in contributing to both engagement and disengagement \cite{engel2023learning}.} In a broader context, these findings challenge the current oversimplified narrative that sees disengagement from problematic ideas as a purely cognitive or information-driven process. The complexity of \textit{exit stories} observed in RQ2 may suggest that disengagement is not just a rational process of belief correction, but also an emotional and identity-driven journey. These findings also have implications for intervention design which are discussed below. 

\subsection{Disengagement from problematic groups vs. established structures}
Overall, the results of RQ2 suggest that disengagement from problematic ideological groups—such as conspiracy theory communities or the manosphere—is qualitatively different from exits in political groups. This distinction emerges both from the content of the reasons cited and the emotional intensity and framing of exit narratives.

While exposure to counter narratives and moral misalignment is cited prominently in both conspiracy theory and political groups, negative mental health, negative self-image heavily underscore exits from conspiracy theory, similar to manosphere groups. In contrast to the exit stories from political groups or religious groups, exits from conspiracy theory and manosphere are often portrayed not just as a change in belief system, but as something corrosive to the self, needing emotional or social repair. Political exits tend to focus on ideological disillusionment, unmet expectations, or shifting worldviews, and are more likely to describe switching sides than full disengagement from political identity altogether. Conversely, exiting problematic ideological communities involves a different type of work—more akin to identity repair or trauma recovery—compared to the more deliberative, value-aligned reorientations seen in religious or political exits. These differences underscore the need for CSCW researchers and designers to approach disengagement from harmful communities not simply as information correction, but as a deeply emotional and moral process requiring different types of support, including recovery-oriented infrastructure, peer validation, and care work. The next sections deliberate more on implications of these findings.

\subsection{Implications for designing interventions against problematic ideologies}
Scholarly research 
% on conspiracy theories 
popularly frames conspiracy theories as primarily ``information problems'', focusing on how misinformation and radical content spread through online platforms. For example, a survey on conspiracy theory research indicates that 72.3\% of all studies examining conspiracy theories \kedit{online} focus on the concepts of misinformation, disinformation, fake news, or rumors \cite{mahl2023conspiracy}. 
This leads to viewing conspiracy radicalization as a byproduct of exposure to false information and ideological content, implying that correcting information or behavior could lead to disengagement. Recent approaches to counter-conspiratorial thinking also reflect this information-centric approach \cite{o2023efficacy}. For example, many interventions focus on inoculating users against conspiracy theories solely through information integrity measures \cite{compton2021inoculation,jolley2017prevention,banas2013inducing,costello2024durably}.  

The results presented in this study, along with recent qualitative works \cite{engel2023learning,xiao2021sensemaking}, 
% however, 
provide necessary insights into the process of disengagement obtained from the lived experiences of people who have already gone through the process of disengagement. Specifically, this paper provides empirical evidence that while counter-exposure is important, the process of disengagement is also influenced by emotions, moral alignment, social networks, and individual identity. 

In the future, research around intervention design can consider aspects other than information for bridging the dialogue with 
individuals engaged with problematic ideologies. For example, researchers can contribute to the design of online spaces as alternatives to 
% conspiracy theory 
groups that offer the same sense of belonging without endorsing 
% conspiratorial 
\kedit{harmful} beliefs. Moreover, future research can also focus on offering positive, inclusive narratives for individuals whose sense of self is tied to conspiracy belief systems. \kedit{For example, g}iven the prominence of moral conflict in disengaging from conspiracy theories, scholars can also focus on designing interventions that, instead of dismissing anxiety around government overreach or corporate corruption, offer pathways for legitimate activism or social justice engagement \cite{douglas2019understanding}. Many accounts in the exit communities also mention falling back to conspiracy theorizing or red pill ideology because of the uncertainty created by world events. Interventions can also focus on building emotional resilience through mental health resources. The next subsections explains the need of mental health support in exit communities in more detail. 

\subsection{Implications in designing mental health support for exit communities}
The results from the analysis of \textit{exit stories} from conspiracy theory, manosphere, and religious communities reveal a critical need for mental health resources in exit communities. Across multiple categories, authors frequently mention the emotional and psychological toll of being involved in such belief systems, with mental health deterioration often cited as a driving factor in their disengagement. In particular, the significant proportion of authors (40\% in conspiracy theories, 58\% in manosphere) discussing negative mental health impacts—such as anxiety, depression, and even suicidal ideation—illustrates how these belief systems can erode emotional well-being.

Many stories of exits from the manosphere reflect how prolonged exposure to red pill and incel ideologies lead to self-loathing, anger, and low self-esteem. Authors also describe hitting ``rock bottom'' before seeking to disengage, highlighting the long-term emotional damage caused by internalizing toxic ideals of masculinity. Similarly, conspiracy theory \textit{exit stories} describe feeling mentally exhausted or drained by the constant exposure to paranoid and negative worldviews that isolate people from their surroundings. Mental health challenges are, in these cases, both a cause and a consequence of engagement with these communities \cite{engel2023learning}.

These findings indicate that mental health interventions are essential for those transitioning out of toxic ideologies. Exit communities, while providing valuable peer support, may not be equipped to handle the emotional complexity or the psychological burden carried by individuals who are disengaging. While forums and subreddits offer spaces to share stories, professional mental health resources—such as counseling, therapy, or support groups—are needed to address deeper emotional scars.

The research community can advance efforts in this direction by designing online spaces with easy, secure, and reliable access to mental health resources. AI-assisted counseling where conversational agents are designed to provide emotional support 
% can 
\kedit{have the potential to} offer valuable help where authors can anonymously share challenges in their recovery. Overall, this study calls for further research on exit communities to offer safe spaces for populations recovering from ideological bounds.  

Importantly, this paper does not advocate for automated interventions or AI-driven deconversion strategies. The goal is not to suggest that AI systems should "disengage" people from communities, but rather to emphasize the importance of understanding disengagement as a complex, user-led process. Any intervention design—whether platform-level, peer-based, or algorithmically mediated—should be grounded in support, care, and user autonomy, not coercion or belief correction. The findings highlight the need for multi-pronged, context-sensitive approaches that recognize disengagement as a gradual and emotionally demanding transition, not a binary decision point.

\subsection{Ethical considerations}
Stories investigated in this paper contain deeply personal accounts shared voluntarily and publicly by Reddit users. \kedit{It is possible that authors are unaware their disclosures are being studied and shared. Further, some} stories shared, especially in the religious communities, contained references to authors changing their identities to manage the consequences of leaving their religions or cults. For this reason, all references to specific religious groups have been removed from the quotes presented in the paper. Further, to preserve the identities of the authors, many quotes cited in the paper are obfuscated with synonymous language \cite{van2020fluids}, and no usernames are mentioned. 

Although sharing datasets and models can support reproducibility and downstream research, the dataset used in this study cannot be released publicly due to ethical and TOS constraints. The model weights supporting code will be released on GitHub upon the acceptance of this paper.

\section{Limitations and Future Directions}
Below I outline the limitations of the current paper and describe how they can shape the future research. First, all accounts of exit analyzed in this paper are limited to individuals who \emph{chose} to share their story and seek social support on Reddit. This limits the applicability of these findings to populations that participate in exit discussions on online forums. Second, exit reason annotations used in RQ2 were conducted by a single coder. This may introduce bias and subjectivity in the RQ2 results. Future refinements of the exit reasons framework can include development with multi-annotator agreements.
Third, the prompts used for multi-class labeling by GPT-4 were contextualized with the mentions of higher level themes such as ``political groups'' or ``religions''. It is possible that some \textit{exit stories} did not get annotated with any reasons because GPT-4 could not determine whether the particular group being described in the story fell under the context provided in the prompt. For example, some religious groups use terms like ``the sea org'', or use abbreviations where describing their insider-outsider status (e.g, PIMO refers to physically-in-mentally-out). It is possible that such peculiarities were missed during the labeling. A better approach could be to provide even more fine-grained prompts by preprocessing the \textit{exit stories} and extracting specific entities being discussed by the authors. Third, while this study presents analysis where several 
% reasons 
\kedit{motivators} co-exist within \textit{exit stories}, future research can focus on understanding causal relationship between different factors. For instance, it is possible that \kedit{in} \textit{exit stories} mentioning personal circumstances as one of the exit reasons, the personal circumstances actually trigger other affective and cognitive processes. Thus, understanding patterns in causal relationships between exit reasons may provide deeper insight about the \emph{process} of disengagement.

While this study adopts a zero-shot multi-label classification approach using GPT-4 due to the complexity of the label space and the variability in linguistic expression, it does not include experimental comparisons with alternative methods such as few-shot prompting, supervised fine-tuning, or domain adaptation. Recent work in NLP has demonstrated that fine-tuning LLMs on task-specific data can outperform zero-shot setups for nuanced psychological and moral concepts {\cite{kim2024moral}}. However, given the large number of overlapping categories in the exit reasons framework and the variability of user-generated text, fine-tuning was not pursued due to cost, complexity, and the lack of sufficient annotated data for all 12 labels. Future research can explore whether fine-tuning on annotated exit stories may yield better generalization or interpretability across community types.

Finally, the sample likely reflects predominantly Western perspectives, both due to Reddit's user demographics and the language of the analyzed content. Reddit has a large English-speaking user base concentrated in North America, Europe, and Australia, which means the exit stories studied here may be shaped by cultural, political, and religious norms specific to these regions. Future research could examine exit narratives from platforms or regions with different cultural and linguistic dynamics to assess cross-cultural differences in disengagement.

\section{Conclusion}
In this paper, I provide a large-scale thematic understanding of various reasons why people leave 
% their 
\kedit{belief and value-driven} communities. Specifically by focusing on five types of \textit{exit stories}---
% political, religious, 
\kedit{politics, religion,} conspiracy theory, manosphere, and lifestyle---I contribute a theory-guided and data-driven understanding of how exits from different types of communities are qualitatively different than each other. I further highlight how information and cognitive processes are not the only mediating factors while disengaging from problematic communities and how disengagement itself is a complex multifaceted process. The results in this work provide a strong empirical foundation for thinking about interventions that focus on improving mental resistance and fostering safe online environments for recovering populations. 

\bibliographystyle{ACM-Reference-Format}
\bibliography{sample-base}

\appendix

\received{October 2024}
\received[revised]{April 2025}
\received[accepted]{August 2025}

\end{document}